# Single-photon emission in the near infrared from diamond colour centre

E Wu[a,b], V. Jacques[a], F. Treussart[a,1], H. Zeng[b], P. Grangier[c] and J.-F. Roch[a]

[a] Laboratoire de Photonique Quantique et Moléculaire, UMR CNRS 8537, ENS de Cachan, 61 avenue du Président Wilson, 94235 Cachan Cedex, France

[b] Key Laboratory of Optical and Magnetic Resonance Spectroscopy, East China Normal University, 3663 Zhongshan Road North, Shanghai 200062, P. R. China

[c] Laboratoire Charles Fabry de l'Institut d'Optique, UMR CNRS 8501, BP 147, 91403 Orsay Cedex, France

Optically active colour centres based on Nickel-Nitrogen impurities are observed in natural diamond under continuous-wave excitation. The spectral analysis shows that the single emitters have a narrow-band emission in the near infrared, around 780 nm, which is almost entirely concentrated in the zero phonon line even at room temperature. The colour centre excited-state lifetime is as short as 2 ns, and the photoluminescence is linear polarized. These striking features pave the way to the realization of a triggered single photon source based on this colour centre emission well suited for open-air single-photon Quantum Key Distribution operating in day-light conditions.

---

[1] Corresponding author. Tel.:+33-1-47407555; Fax: +33-1-47402465
E-mail address: francois.treussart@physique.ens-cachan.fr (F. Treussart)

1. Introduction

Single photons on demand can be efficiently produced by suited pulsed excitation of a single emitting dipole, such as a single atom, molecule, quantum dot and diamond colour centre [1]. Since it takes a single emitting dipole a finite time to go through a full cycle of excitation-emission-reexcitation before it emits a second photon, a sufficiently short and intense excitation pulse will enable the single dipole to emit only one photon per excitation pulse [2]. Among various solid-state sources, single defects in diamond are one of the potential candidates due to their unsurpassed efficiency and photostability. The Nitrogen-Vacancy (NV) colour centre is one of the best known optically active defects in diamond [3-6]. It consists of a substitionnal nitrogen atom (N) and a vacancy (V) in an adjacent lattice site, emitting around 670 nm with a zero phonon line (ZPL) at 637 nm. A reliable triggered single-photon source was recently built, based on the pulsed, optically excited photoluminescence of a single NV colour centre in diamond nanocrystal [7]. This single-photon source was applied to the observation of single-photon wavefront-splitting interference [8] and the realization of open-air quantum key distribution (QKD) experiments [9,10]. However, in the latter application, broadband emission of the NV colour centre (FWHM≈70 nm at room temperature) precludes day-light operation of the QKD setup due to difficulties in filtering the transmitted single-photons from day-light background.

Recently, the Nickel-Nitrogen NE8 impurities in the diamond arouse the researchers' interest for their intense narrow emission band in the near infrared [11,12]. In this article, we report on the observation of similar diamond colour centres also tentatively associated to Nickel-Nitrogen impurities. Photoluminescence of these single emitters shows perfect photostability at room temperature. Compared to NV colour centre emission, it has several striking features. I. A narrow band emission around 780 nm which is almost entirely concentrated in a ZPL even at room temperature. II. A short photon emission lifetime, around 2 ns. III. Linearly polarized photoluminescence from the single defect.

2. Experimental setup

The experimental set-up is sketched in Fig.1. Confocal microscopy is employed to select the single defects in the sample. The sample is a Type IIa natural bulk diamond wafer 3 × 3 mm$^2$ in size and 1 mm in thickness (Element 6, Netherlands). A commercially available laser diode emitting at 687 nm provides the CW excitation source. The beam is focused on the sample about 4 μm below its surface using a microscope objective with a magnification of 100 and a numerical aperture of 0.95, yielding a spot size of about 1 μm (FWHM). Photoluminescence from the excited defects in the sample is collected by the same microscope objective and sent to the detection setup after proper frequency and spatial filtering. The detection system includes a home-built spectrograph (spectral resolution 1 nm) imaging the spectrum on a cooled CCD array and a Hanbury Brown and Twiss setup (HBT) which consists of a beamsplitter and two silicon avalanche photodiodes in single-photon counting regime (APD, Perkin Elmer). The spectral analysis gives the information on the identification of the various impurities in the diamond sample. The HBT time coincident setup provides fluorescence intensity autocorrelation measurements from which emission from a single emitter is readily identified. For this purpose, the two photodiodes are connected to the start and stop inputs of a time-to-amplitude converter (TAC). The TAC output is stored in a multi-channel analyser (MCA) which yields the histogram of time-delay τ between two consecutively detected photons. Considering our detection efficiency of a few percent, this histogram is, on short time-scale, a good approximate of the intensity *I(t)* second-order correlation $g^{(2)}(\tau)$ [13], defined by:

$$g^{(2)}(\tau) \equiv \frac{\langle I(t)I(t+\tau)\rangle}{\langle I(t)\rangle^2} \tag{1}$$

3. Results and discussion

To identify well isolated photoluminescent emitters, we first raster scan the sample using CW excitation. For each photoluminescent spot, the unicity of the emitter is then checked by observation of antibunching in the photoluminescence intensity. Since after the emission of a first photon it takes a finite time for a single emitter to be excited again and then spontaneously emits a second photon, the antibunching effect appears as a dip around zero delay $\tau = 0$ in the normalized intensity autocorrelation function $g^{(2)}(\tau)$.

For each isolated emitter we simultaneously record the spectrum of the emitted light and the histogram of time delays between the consecutively detected photons. Fig.2 shows the spectrum of the light from such an emitter. The narrow peak at 782 nm (about 2 nm FWHM) corresponds to ZPL of Nickel-Nitrogen related impurities [14,15], and the sharp peak at 757 nm is the one phonon Raman scattering line of the diamond lattice (1332 $cm^{-1}$) associated to 687 nm excitation wavelength. Note the remarkable property that photoluminescence is concentrated in the ZPL even at room temperature. The ZPL integrated intensity corresponds to 68% of the full spectrum area, a value much higher than that of NV colour centre.

The histogram of time delays is recorded with the HBT setup after filtering of the ZPL with a narrow band-pass filter having its inclination relative to the beam normal incidence tuned for maximal transmission. Since the signal to background ratio is quite large (60:1), we neglect the residual background light effect on the $g^{(2)}$ measurement. Fig.3(a) shows the intensity autocorrelation function $g^{(2)}(\tau)$ of the light from the same emitter as the one studied in Fig.2. A clear dip of the autocorrelation function is observed at zero delay with $g^{(2)}(0) = 0.12$, proving that we address a single emitter. However the non-zero $g^{(2)}(0)$ value can not be attributed to background poissonian light, the intensity of which is negligible owing to very high signal to background ratio. This residual value is indeed due to the detection setup time instrumental response function (IRF). This function is measured to be a 0.7 ns FWHM gaussian function, by autocorrelation of titanium:sapphire laser 150 fs pulses. Taking into account this finite time response, the measured autocorrelation function can be well fitted by the convolution of the IRF with a $g^{(2)}(\tau)$ function going perfectly to zero at $\tau = 0$, and modelled by a symmetrical exponential function in the framework of a three-level model (see Fig.3(a),Ref [5]). Note that value of $g^{(2)}(\tau)$ higher than unit at delays larger than 10 ns, is due to the leakage of the system towards a dark metastable state, inducing photon bunching for the corresponding time-scale. In order to determine the intrinsic excited-state lifetime of the single defect, we measured the characteristic time of the $g^{(2)}$ at different excitation powers. Fit of this time in the framework of a three-level model yields a value of the excited-state lifetime of the single colour centre of about 2 ns, obtained by extrapolating the fit at zero excitation power (Fig.3(b)).

We also investigated the polarization properties of the single colour centre relative to absorption and emission of light. We monitored the photoluminescence intensity while rotating the excitation laser linear polarization angle as shown of Fig.4. The curve contrast of 96% proves that the single colour centre behaves like a dipole relative to absorption of light. Using quarter wave plate method, we studied the emitted light polarization properties and observed that light is perfectly elliptically polarized with an aspect ratio of 0.25. This measurement indicates that the single defect also behaves like an emitting dipole. The linear polarized light becomes elliptical after propagation though optics including high numerical aperture objective and dichroic mirror, both known for modifying the polarization state of light.

Let us finally point out that the single colour defect is a quite strong emitter. For the maximum available excitation power of 9 mW, an overall counting rate of 300 kcounts/s is obtained corresponding to an excited-state population occupancy of about 60%.

3. Conclusion

In conclusion, we have studied photoluminescence properties at the single emitter level of optically active colour centres based on Nickel-Nitrogen impurities in natural diamond samples. Under CW excitation at room temperature, these colour centres reveal narrow band emission abound 782 nm, the short photon emission lifetime about 2 ns, and fully polarized photoluminescence. Thanks to narrow spectral emission, it will be easy to pick up the useful photo luminescence while removing stray light. Thanks to short photon emission life time, implementation of the single-photon source in QKD setup will be compatible with narrow time-window analysis. This will limit the effect of photodetection dark counts and will lead to an increase of the limit of propagation distance compatible with perfectly secure communications [10,16]. We can also take advantage of the perfectly polarized light which can easily be turned into linear polarization with a properly oriented quarter wave plate, virtually without any loss. Finally, not that the emission wavelength of the single emitter is both in an open-air and telecom optical fibre communication window.

With all of those striking features, the colour centres based on diamond Nickel-Nitrogen impurity appear to be very good candidates for realizing a stable triggered single-photon source at room temperature for an efficient and practical open-air QKD system.

Acknowledgements

This work was partly supported by "AC Nanosciences" grant from Ministère de la Recherche, by Institut d'Alembert (ENS de Cachan IFR 121) and by Institut Universitaire de France.

Figure Captions

Fig.1 Experimental setup. LD: Laser diode emitting at 687 nm; MM: Mobile mirror fixed on a piezoelectric transducer providing orthogonal angular displacements; O: Microscope objective; DM: Dichroic mirror; PH: Pinhole (100 μm diameter); F1: Interference filter transmitting λ >740 nm and removing the remaining light at the excitation wavelength; BS: non-polarized beam splitter; F2: Band-pass filter (FWHM ≈ 10 nm); APD: Silicon avalanche photodiode in photon counting regime; TAC: Time-to-amplitude converter.

Fig.2 Photoluminescence spectrum from a single Nickel-Nitrogen impurity in the diamond sample. The sharp intense line at a wavelength of 757 nm corresponds to the one phonon Raman scattering of the diamond matrix for excitation at 687 nm.

Fig.3 (a) Normalized correlation function $g^{(2)}(\tau)$ measured with CW excitation power of 0.68 mW. The experimental data are shown as dots, while the solid line represents a fit by a convolution of a symmetrical exponential function (characteristic time $\tau_c$) with IRF. (b) characteristic time $\tau_c$ of the $g^{(2)}$ function for various excitation powers. Solid line: fit in the framework of a three-level model. Extrapolated value of the fit at zero pump power yields the excited-state lifetime of about 2 ns.

Fig.4 Photoluminescence (in kcounts/s) vs. excitation linear polarization angle for a single Nickel-Nitrogen defect in diamond. Solid line: fit by a Malus-type law, yielding a contrast of 96%.

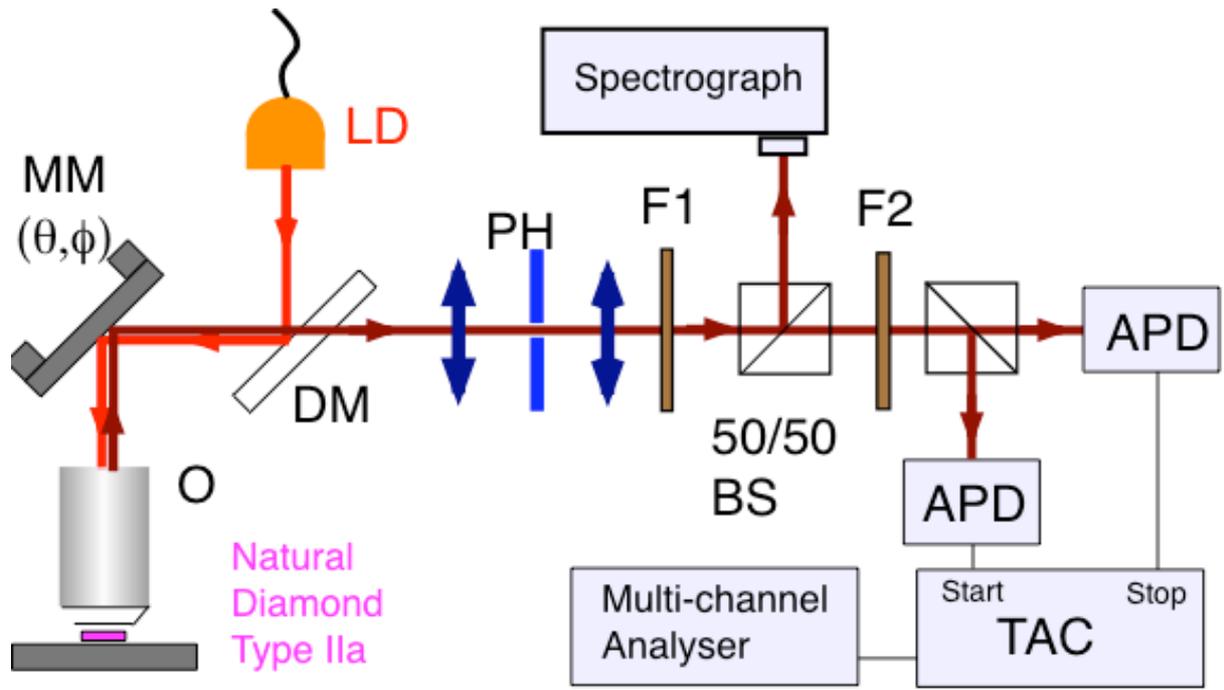

Fig.1 Wu et al

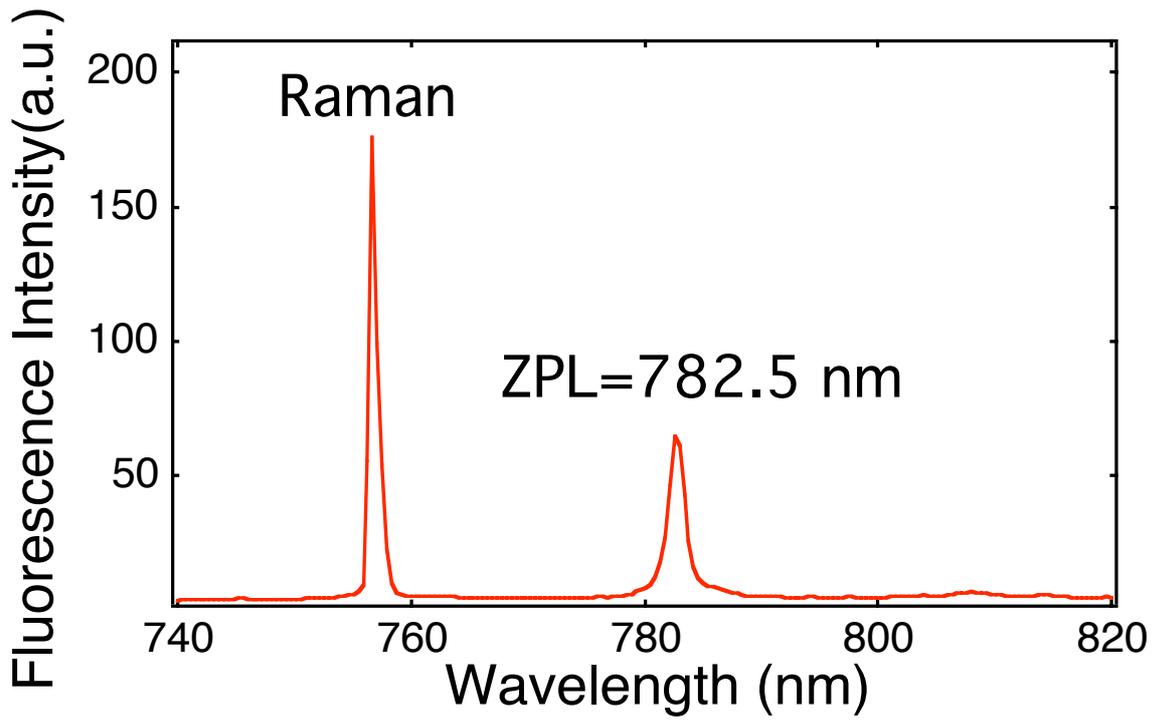

Fig.2 Wu et al

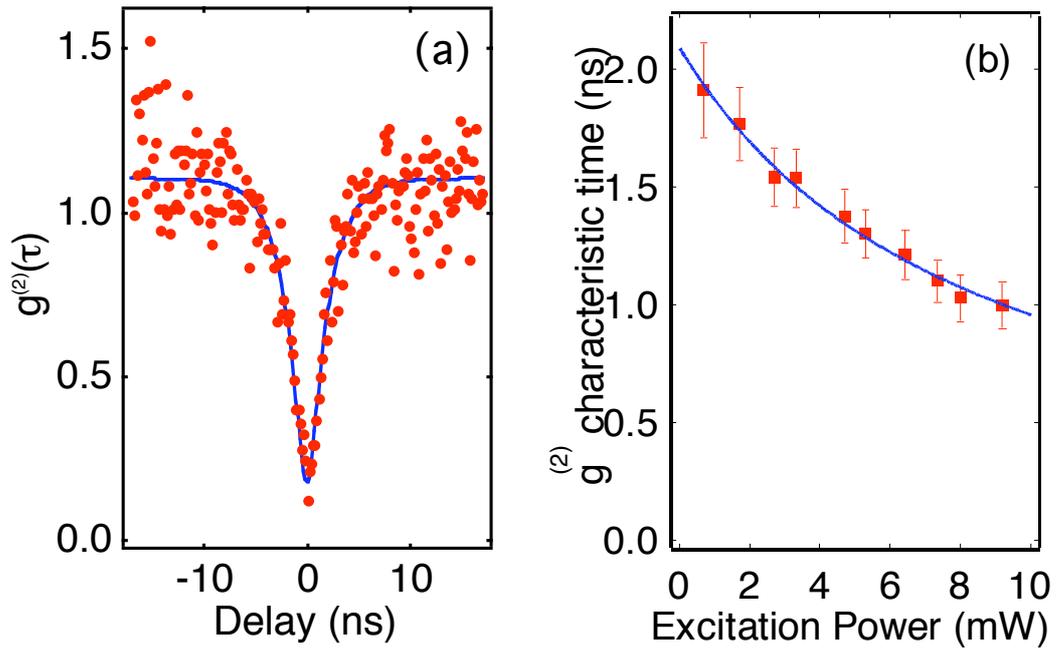

Fig.3 Wu et al

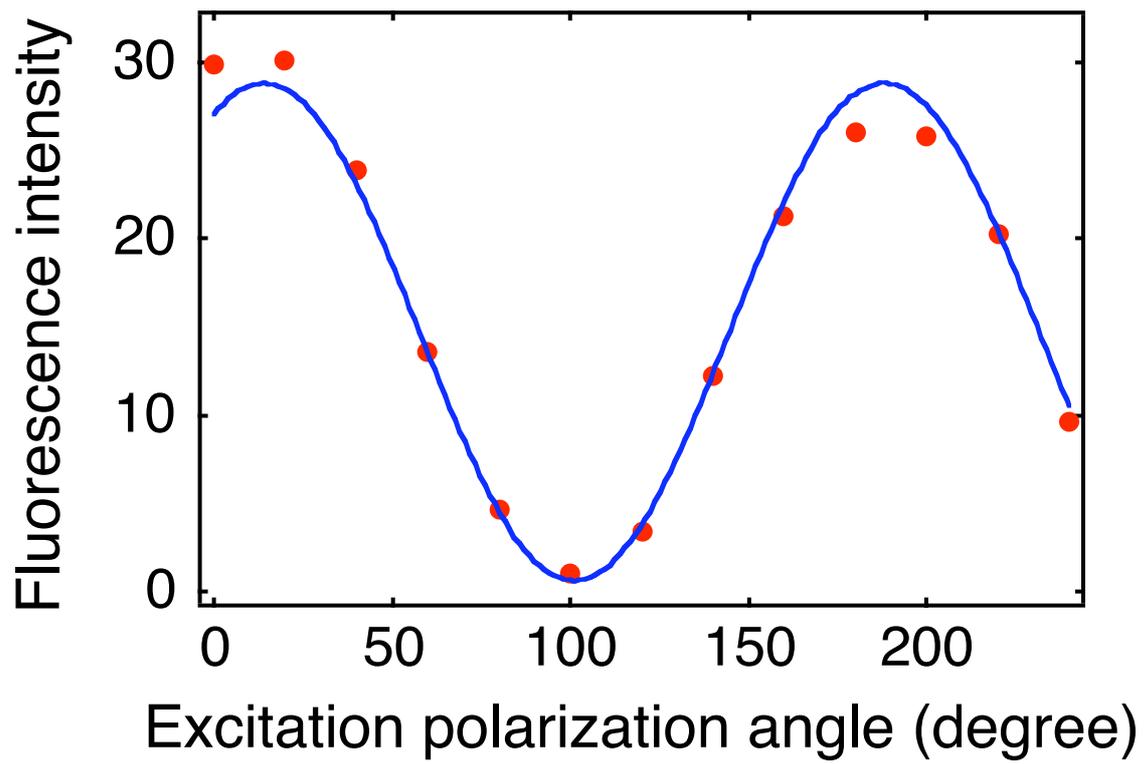

Fig.4 Wu et al